\documentstyle[spie]{article} 
\input{psfig}   

\def\la{\hbox{\rlap{\raise.3ex\hbox{$<$}}\lower.8ex\hbox{$\sim$}\ }}
\def\ga{\hbox{\rlap{\raise.3ex\hbox{$>$}}\lower.8ex\hbox{$\sim$}\ }}

\title{Construction and Testing of a Pixellated CZT Detector and
Shield for a
Hard X-ray Astronomy Balloon Flight} 

\author{P. F. Bloser, T. Narita, J. A. Jenkins, and J. E. Grindlay
\skiplinehalf 
Harvard-Smithsonian Center for Astrophysics, 60 Garden St., 
Cambridge, MA 02138, USA
}

\authorinfo{Further author information: (Send correspondence to
P. Bloser)\\P.B.: E-mail: pbloser@cfa.harvard.edu}

\begin{document} 
\maketitle 

\begin{abstract}
We report on the construction and laboratory testing of
pixellated CZT detectors mounted in a
flip-chip, tiled fashion and read out by an ASIC, as required for
proposed hard X-ray astronomy missions.  Two 10 mm $\times$ 10 mm
$\times$ 5 mm 
detectors were fabricated, one out of standard eV Products
high-pressure Bridgman  CZT and one out of IMARAD horizontal Bridgman
CZT.  Each was fashioned with a 4 $\times$ 4 array of gold pixels
on a 2.5 mm pitch with a  surrounding guard ring.  The detectors
were mounted side by side on a  carrier card, such that the pixel
pitch was preserved, and read out by a 32-channel VA-TA ASIC from
IDE AS Corp.
controlled by a PC/104 single-board computer.  A passive
shield/collimator surrounded by plastic scintillator encloses the
detectors on five sides and provides a $\sim 40^{\circ}$ field of view.  
Thus
this experiment tests key techniques required for future hard X-ray
survey instruments.  The experiment was taken to Ft Sumner, NM in May
2000 in preparation for a scientific balloon flight aboard the joint
Harvard-MSFC EXITE2/HERO payload.  Although we did not receive a
flight opportunity, and are currently scheduled to fly in September
2000, we present our calibration data in the flight configuration
together with data analysis techniques and simulations of the expected
flight background spectrum.
\end{abstract}



\section{INTRODUCTION}
\label{sec:intro}  

It has been more than twenty years since the last all-sky hard X-ray
($> 20$ keV) survey was performed by the HEAO A-4 instrument\cite{levine}.
A new survey, performed with a high-sensitivity imaging instrument, is
urgently needed.  Cadmium Zinc Telluride (CZT) will almost certainly be
the detector material that finally makes this new survey a reality.  CZT
offers far better energy resolution than scintillator detectors such as 
NaI and CsI, and the use of pixel or strip electrode readouts allows far
better spatial resolution.  Response up to 600 keV is possible with 
moderate thicknesses (5 mm) due to the high stopping power of CZT, and
no cryogenic cooling is required due to its wide bandgap.  Based on these
advantages, CZT has already
been selected for its first space-based application as a gamma-ray burst
monitor on the SWIFT mission\cite{barthelmy}.  Our Harvard group has been 
developing more
advanced CZT detectors for astronomy applications, motivated by the
requirements of a wide-field, all-sky survey telescope operating between
$\sim 10$ and 600 
keV\cite{bloser98mrs,bloser98,narita98,bloser99,narita99,narita2000}.
Specifically, we have focused on the EXIST\cite{grindlay2000} or  
EXIST-LITE\cite{grindlay98} concepts as our baseline.

Before such instruments can be constructed, several technical issues remain
to be addressed.  Only the coded-aperture technique allows imaging between
100 keV and 600 keV, and this requires large-area, position-sensitive 
detectors that are subject to large background levels.  
For a sensitivity of $\sim 0.05$ mCrab, as baselined for EXIST on the
International Space Station, approximately 8 m$^2$
of CZT are required\cite{grindlay2000}.   
Large fields of view
($40^{\circ}$ for each module of EXIST) are needed to conduct a survey over 
the entire sky, further increasing background and requiring
large (1--2 mm) pixels to avoid projection effects in thick detectors.  
Large pixels have increased leakage current noise, and this must be reduced
to allow high bias voltages for complete charge collection.  The highest
resistivity CZT available is high-pressure Bridgman (HPB) material.  While 
detectors made of HPB CZT have low leakage current, they can at present 
only be made 10 mm $\times$ 10 mm in size with a reasonable yield, making 
the construction of 
square meter detector planes a technical challenge.  Recently IMARAD 
Imaging Systems began producing CZT using a modified horizontal Bridgman
(HB) process\cite{cheuvart90} which allows the growth of larger crystals 
(40 mm $\times$ 40 mm) at higher yield and thus lower cost.  
IMARAD CZT also appears to be more uniform than HPB
material\cite{narita99} which, if confirmed, makes it more suitable
for large area arrays since inter-pixel calibration is less demanding.
The HB CZT 
has a lower resistivity,
however, and thus higher leakage current.  We have found that appropriate
contact material can reduce the leakage current in HB detectors to levels
comparable to HPB material\cite{narita99}; in particular, simple gold 
electrodes on IMARAD CZT act as blocking contacts and significantly reduce
leakage current noise\cite{narita2000}.

Thus the main challenges to be addressed at present are: find a way to
package small detectors that will allow them to be tiled into large arrays
with a minimum of dead space; find appropriate contact materials to reduce
leakage current enough to allow the use of lower-cost material; read out 
and process signals from thousands of pixels using low-noise ASICs that do
not take up additional space; and find optimal shielding techniques to
minimize the background in CZT in the high-radiation environment of low
earth orbit with a minimum of mass and complexity.  We have constructed
a balloon experiment to test methods of meeting these challenges and to 
measure the spectrum and uniformity of the background generated in an 
imaging CZT detector.  The motivation and basic design of this experiment
have been described previously\cite{bloser99}.  The completed instrument 
was taken to Ft. Sumner, NM in May 2000 for a balloon flight on the joint
Harvard-MSFC EXITE2/HERO payload.  We did not receive a flight opportunity
and are currently scheduled to fly in September 2000.  In this paper,
therefore, we describe the completed instrument and its performance in 
detail and present preliminary simulations of its response and the 
flight background expected.

\section{Description of the Imaging CZT Experiment}
\label{sec:description}

The heart of the Harvard imaging CZT experiment is the tiled
``array'' of flip-chip-mounted pixellated detectors.  This array
consists of two moderately thick (10 mm $\times$ 10 mm $\times$ 5 mm)
crystals, 
one made of HPB CZT provided by eV Products and one made of HB CZT
provided by IMARAD.  These crystals were fashioned into detectors with
gold contacts by RMD, Inc. using a direct shadow mask and evaporator
technique\cite{narita2000}.  Gold contacts were selected for both 
crystals in order to make a well-understood ohmic detector out of the
eV Products material and because we have found that gold contacts on
IMARAD material act as blocking contacts and greatly reduce leakage
current\cite{narita2000}.  
\begin{figure}
\begin{center}
\begin{tabular}{c}
\psfig{figure=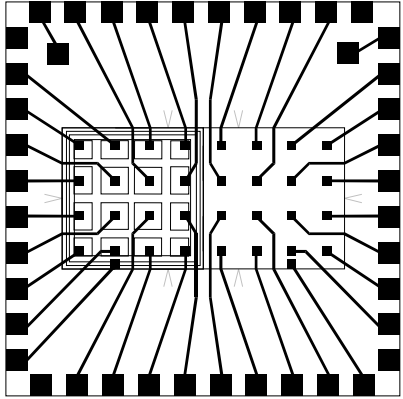,height=6cm} 
\end{tabular}
\end{center}
\caption[initial] 
{ \label{fig:layout}        
The Harvard flip-chip detector layout and carrier card design.  Each
CZT detector has a $4 \times 
4$ array of pixels surrounded by a guard ring.  The edge and corner
pixels are made smaller to allow the pixel pitch to remain constant
between detectors.  The detectors are mounted side-by-side on a
ceramic carrier card with gold pixel pads and traces carrying the
signals to pins around the edge.
} 
\end{figure}
The pixel layout and tiling configuration
is shown in Figure~\ref{fig:layout}, along with the design of the
detector carrier card.  Each detector was made with
a $4 \times 4$ 
array of pixels on a 2.5 mm pitch with 150 $\mu$m gaps, surrounded by
a guard ring.  The inner pixels were made 2.35 mm across, while the
edge and corner pixels were made slightly smaller so that the 2.5 mm
pitch could be preserved when the detectors were mounted
side-by-side (Figure~\ref{fig:layout}).  

Both detectors were tested using a pogo pin and eV
Products preamp prior to flip-chip mounting.  
\begin{figure}
\begin{center}
\begin{tabular}{c}
\psfig{figure=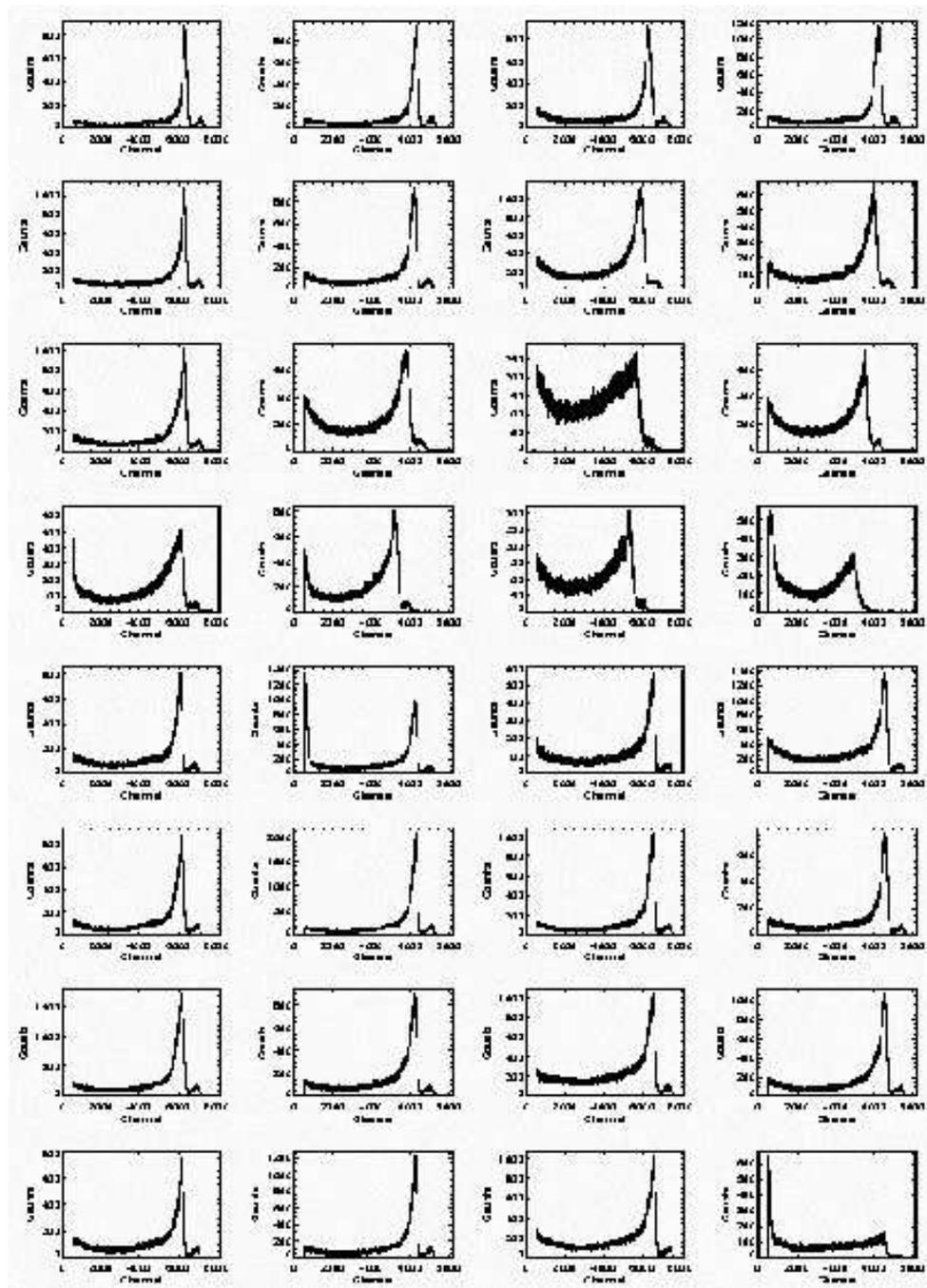,height=18cm} 
\end{tabular}
\end{center}
\caption[initial] 
{ \label{fig:preflip}        
$^{57}$Co spectra recorded in the IMARAD detector (top 4 $\times$ 4 spectra)
and eV Products detector (bottom 4 $\times$ 4 spectra) prior to
flip-chip mounting.  Each spectrum was recorded individually using a
pogo pin and eV Products preamp.  The lower right corners of both
detectors show bad spectra due to material defects.
} 
\end{figure}
Figure~\ref{fig:preflip} shows the 32 spectra recorded in this
manner using a $^{57}$Co source.  The top 16 spectra are from the
IMARAD detector and the bottom 16 from the eV Products detector; the
plots are arranged in the order of the pixels when both detectors are
mounted.  Data were taken from each pixel
individually; the integration times were not identical and the
detector had to be repositioned between each run, so the illumination 
was not uniform.  The spectra are generally quite good, except for areas of
material defects in the lower right corners of both detectors.  

The two CZT detectors were mounted in a flip-chip fashion on a ceramic
carrier board by HyComp, Inc.  The carrier card was made with a $4
\times 8$ array of gold pixel pads on a 2.5 mm pitch, with traces running
from each pad to pins on the edge (Figure~\ref{fig:layout}).
Additional pads were included to 
ground the guard rings and bring high voltage onto the card.  A gold
wire bond bump ($\sim 100$ $\mu$m across) was attached to each pad and
coined flat.  Conductive 
epoxy was applied to the pads with a stencil and the detectors were
aligned and mounted.  After the conductive epoxy had cured a
non-conductive epoxy underfill was injected under the detectors to
provide mechanical strength.  
\begin{figure}[t] 
\begin{minipage}[t]{3.3in}
\psfig{file=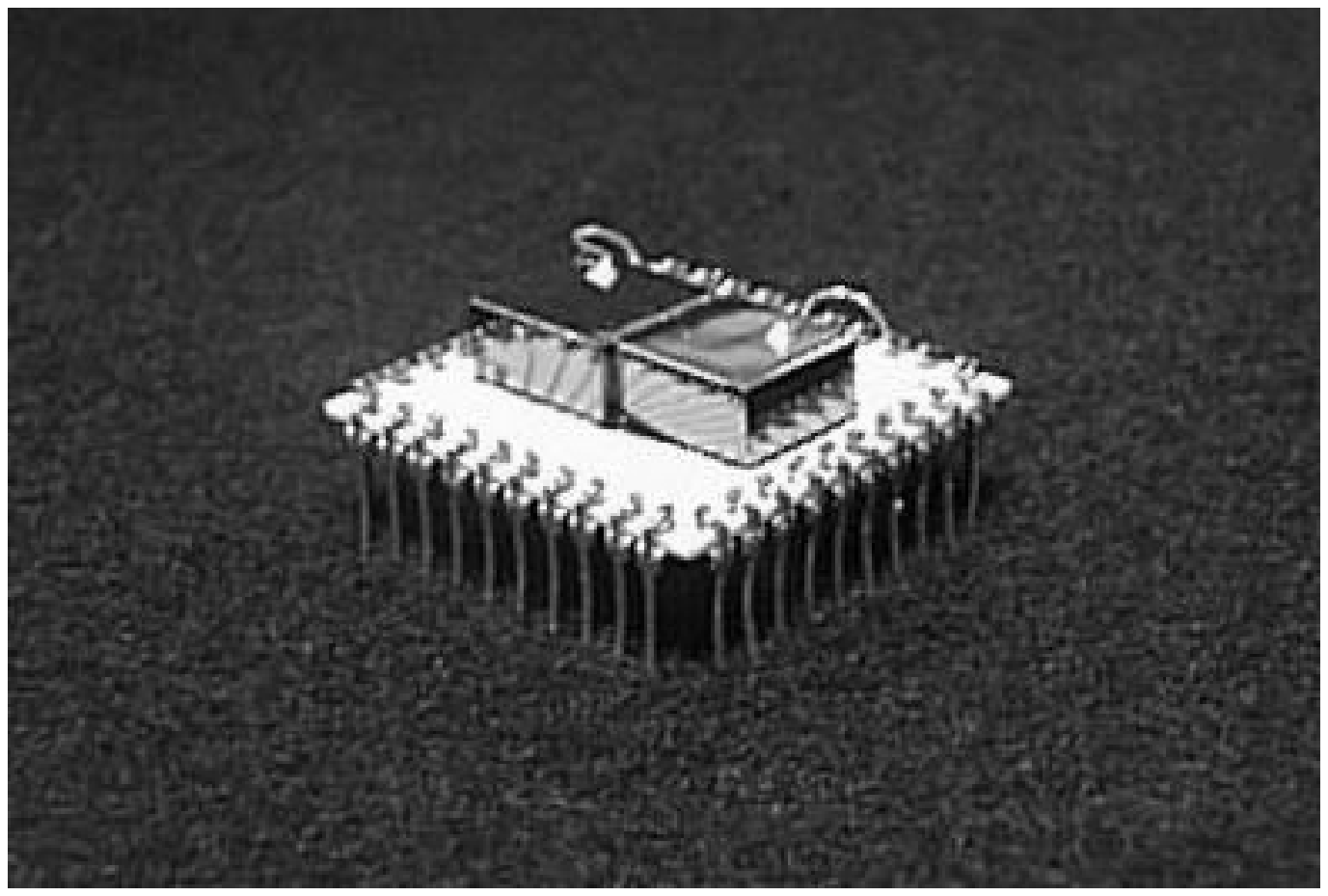,height=2.1in,width=3.2in}
\caption{The two CZT detectors mounted flip-chip fashion on the
ceramic carrier card.  The pixel pitch is maintained across the two
detectors, as required for an imaging detector array.  The thin
wires epoxied to the gold cathodes supply high voltage.}
\label{fig:detectors}
\end{minipage}
\hspace*{0.2in}
\begin{minipage}[t]{3.3in}
\psfig{file=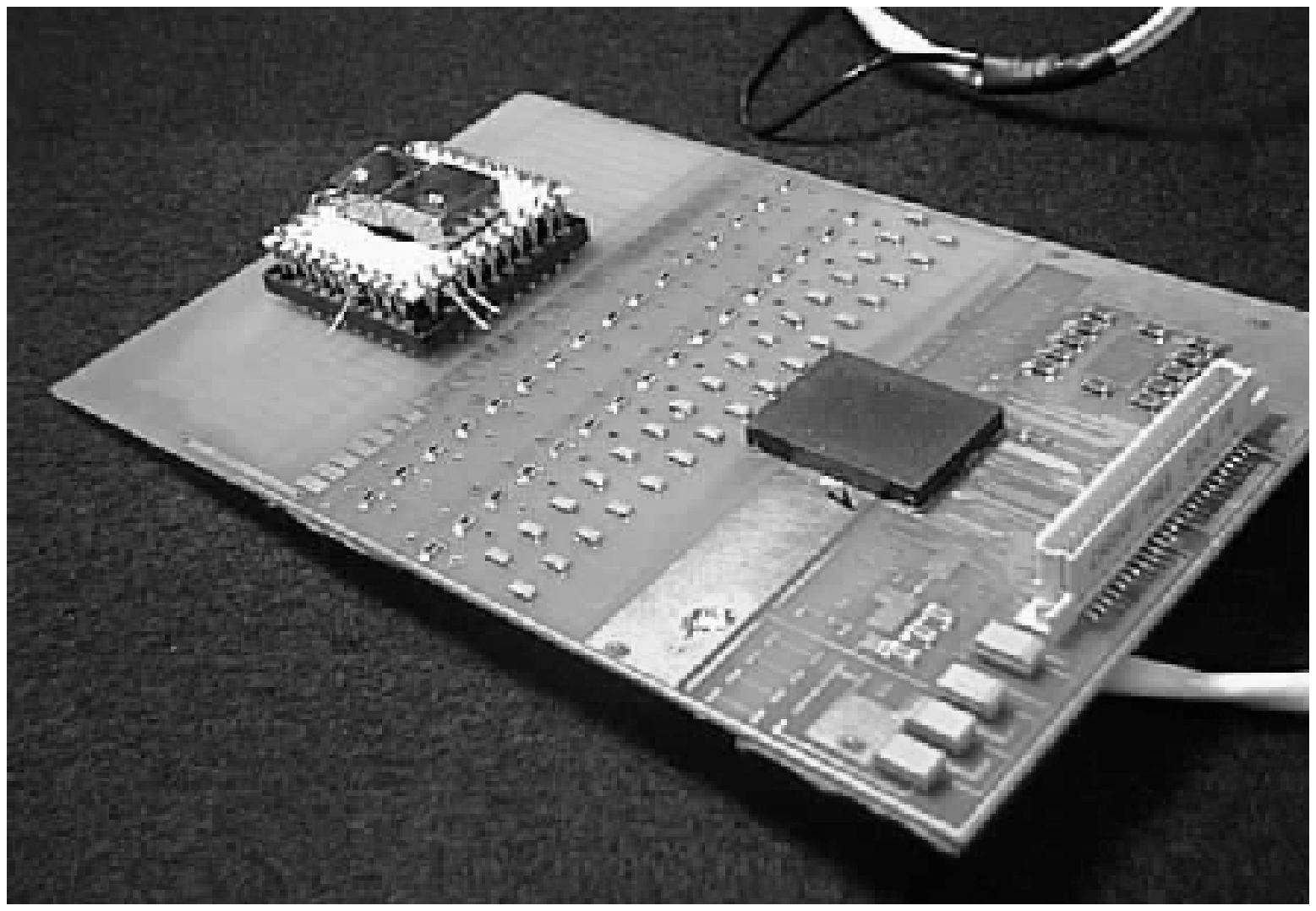,height=2.1in,width=3.2in}
\caption{The custom-designed 32-channel VA-TA ASIC card, supplied by
IDE AS Corp., with the
flip-chip CZT detector plugged 
in.  The VA-TA is mounted under the protective cover on the right
side.  The VA-TA is read out by a commercial DAQ board, which is in
turn controlled by a PC/104 computer. }
\label{fig:vata}
\end{minipage}
\end{figure}
In Figure~\ref{fig:detectors} we show the assembled flip-chip CZT detector
mounted on its carrier card.  Negative high voltage is applied 
via the two thin wires epoxied to the gold cathodes.  

An second identical flip-chip detector was manufactured for us by HyComp
using additional IMARAD and eV Products CZT crystals.  The raw CZT
material was of slightly higher quality in this second detector;
however, after assembly we found that several IMARAD pixels were
partially shorted to the high voltage input.  As this would endanger
the readout 
electronics, we have not proceeded with testing this detector.  We
report here only results for the first detector, which we will fly in
September.  

Both detectors are read out by a self-triggering, 32-channel VA-TA
ASIC supplied by IDE 
AS Corp. in Norway.  The CZT detector card plugs directly into a
custom-designed VA-TA carrier board, as shown in
Figure~\ref{fig:vata}.  The ASIC is read out by a commercial data
acquisition (DAQ) board also supplied by IDE.  This VA-DAQ board is in
turn controlled by a PC/104 single-board computer running DOS software
adapted by us from IDE-supplied Labview code.  This code allows us to
set the trigger threshold and to mask out noisy channels.  The VA-TA
card is a 
preliminary design made for our test balloon flight only; it is not
suitable for our final survey telescope due to its large area and long
lead lengths between detector and ASIC.

\begin{figure}
\begin{center}
\begin{tabular}{c}
\psfig{figure=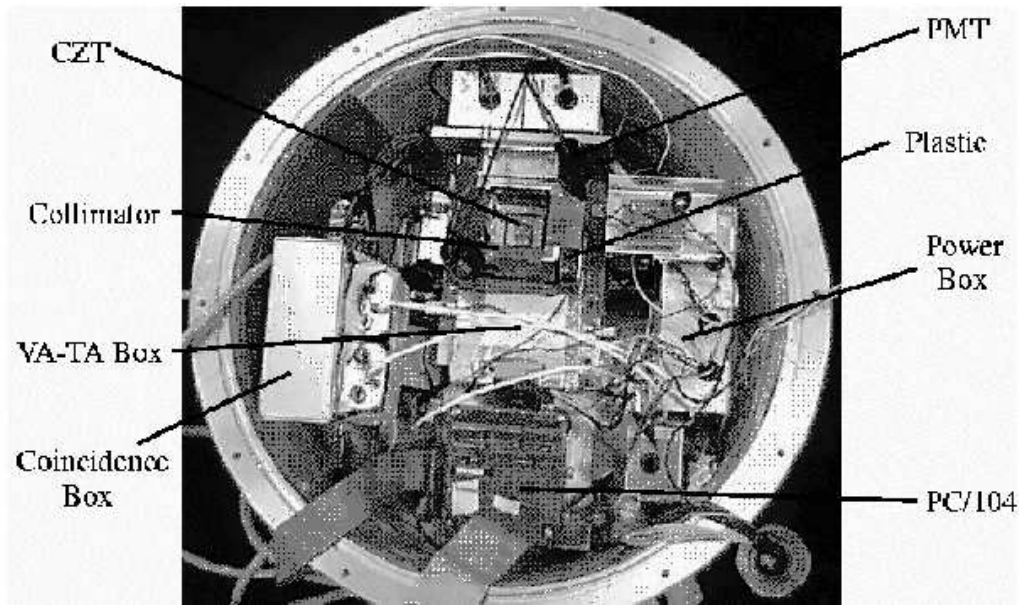,height=8cm}
\end{tabular}
\end{center}
\caption[initial]
{\label{fig:czt2}
The complete Harvard CZT experiment in its pressure vessel, looking
down the collimator at 
the CZT detectors.  Visible are the passive
collimator with its surrounding plastic scintillator and readout PMTs,
the VA-TA shield box, the PC/104 computer, shield
coincidence logic electronics box, and power regulation boxes.  The
experiment is scheduled to fly from Ft. Sumner, NM in September
2000.
}
\end{figure}

The fully-assembled CZT experiment is shown in its pressure vessel in
Figure~\ref{fig:czt2}.  
The detectors are visible slightly above center.  They are surrounded
by a passive Pb/Sn/Cu collimator which is in turn surrounded by
plastic scintillator read out by miniature PMTs (Hamamatsu
R7400U, 1.5 cm diameter).  The collimator
provides a $40^{\circ} \times 40^{\circ}$ field of view, and the plastic
scintillator will veto local gamma-rays produced in the passive
material by cosmic ray interactions.  A similar rear shield, not
visible, sits below the CZT.  (Further details of the shielding
configuration have been described by us previously\cite{bloser99}.)
The VA-TA card is enclosed in a thin aluminum shield box to isolate it
electrically from the rest of the hardware; we have found that the
ASIC is extremely susceptible to ground loop and pickup noise.  At
the bottom of the figure is the PC/104 computer that controls the data
acquisition, reading out all 32 channels for each trigger.  The PC/104
stack includes an analog/digital I/O card (DM5408) that allows the
computer to 
set thresholds for the PMTs, switch the CZT and PMT high voltage on
and off, and monitor coincident CZT and plastic
shield events.  The coincidence electronics are enclosed in the box at
left.  Computer and other electronic power is distributed in the boxes
on the right.  Not visible is the compact high voltage supply (C12N)
provided by EMCO High Voltage Corp.  Normal operating bias for the CZT
detectors is -500 V.  
The ASIC and high voltage supply are powered by isolated batteries,
which we have found to be critical for low-noise operation.  A
temperature sensor is included and read out by the PC/104.  In normal
operating mode the PC/104 collects events for one second, recording
all 32 channels and the plastic shield flags for each event, and
packages them into a buffer with the housekeeping data.  These buffers
are sent out to the main EXITE2/HERO data stream via the serial port.
We are limited by overall balloon system throughput requirements to 50
CZT events per buffer.

\section{CZT Detector Testing and Data Analysis}
\label{sce:testing}

The leakage current was measured as a function of bias voltage for
both detectors for each pixel.  The voltage was varied between -600 V
and +200 V.  It was immediately obvious that three pixels on the IMARAD
detector were shorted to ground.  The pins connected to these pixels
were lifted so as not to connect to the ASIC.  It is not clear why
this happened only on the IMARAD side; we suspect that the non-conductive
underfill epoxy reacted differently with the IMARAD material than with
the eV Products material, perhaps creating a conducting oxide layer
that led to shorts.
The leakage current on the other pixels was 
fairly uniform, with values of $\sim$ 1--2 nA per pixel at -500 V for
both detectors.

Calibration spectra were taken by fully illuminating both detectors
simultaneously using $^{241}$Am and $^{57}$Co sources.
The ASIC trigger threshold was set at $\sim 30$ keV.  We
discovered that two channels on the eV Products detector were dead,
and that six additional IMARAD pixels were too noisy to use
with our 50 cts buffer$^{-1}$ limitation.  These pixels were mostly on
the far side of the IMARAD detector from the VA-TA chips and therefore
had the longest traces on the VA-TA card.  We masked these channels
out.

The spectra in the remaining 21 channels were acceptable, but could be
greatly improved through simple data post-processing.  The spectral
degradation was due to
charge-sharing between pixels for events that occurred near pixel
boundaries.  Since our pixels are quite large, charge is only shared
between two pixels in most cases.  We therefore adopted the following
procedure: 1) We first performed a rough energy calibration on each
channel using the
60 keV line from $^{241}$Am and 122 keV line from $^{57}$Co.  2) For
each event, we found the channel containing the largest pulse height.
The photon presumably interacted in this pixel.  3) We searched the
surrounding 4 ``nearest neighbor'' pixels for the maximum pulse
height.  This is the pixel into which charge may have been shared.  4)
The pulse height from the neighbor pixel is adjusted by the relative
gain and offset and added to the pulse height of the central pixel.
Thus each event is corrected for charge loss due to sharing between
pixels.

\begin{figure}
\begin{center}
\begin{tabular}{lr}
\psfig{figure=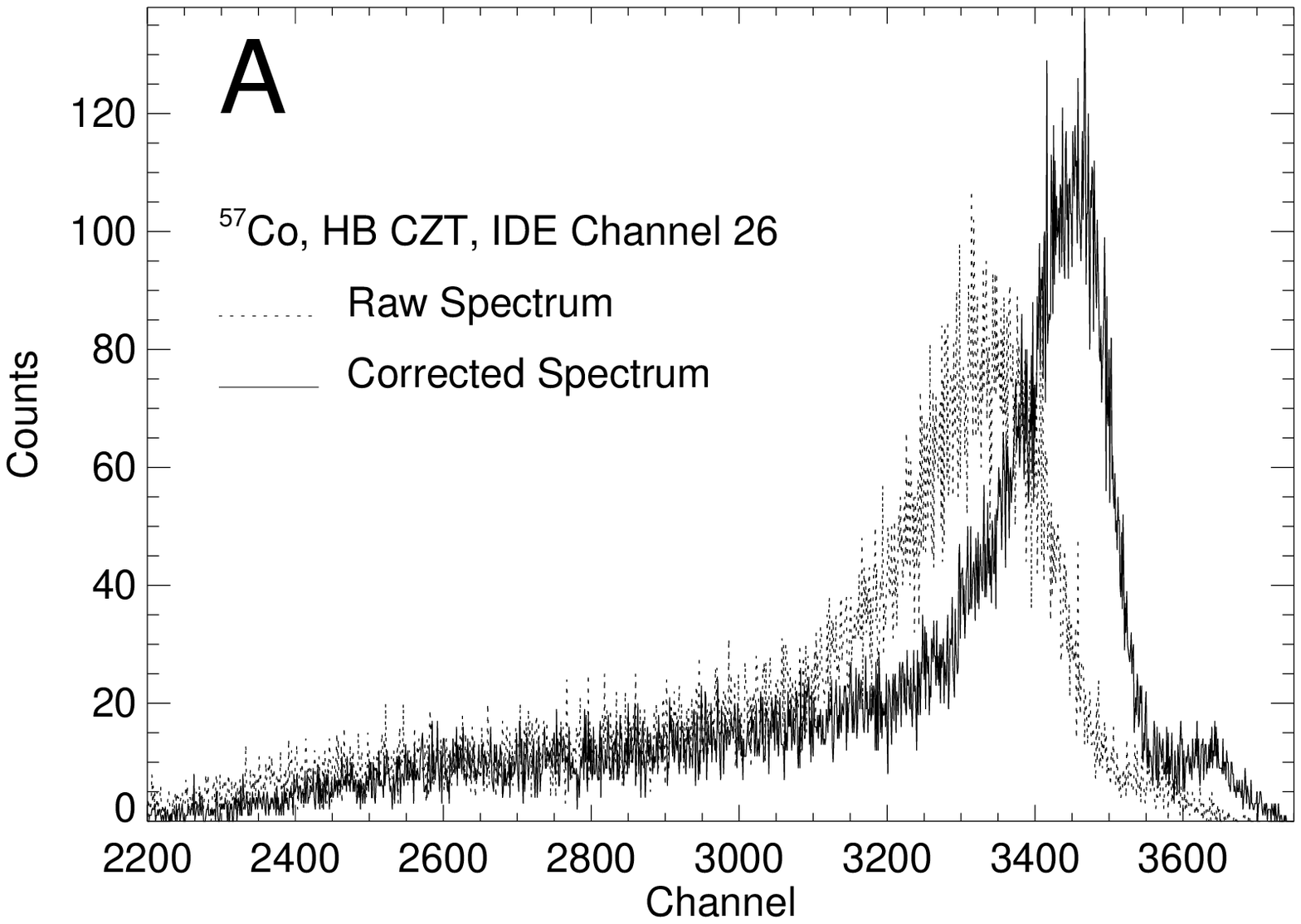,height=2.1in,width=3.2in}
& \psfig{figure=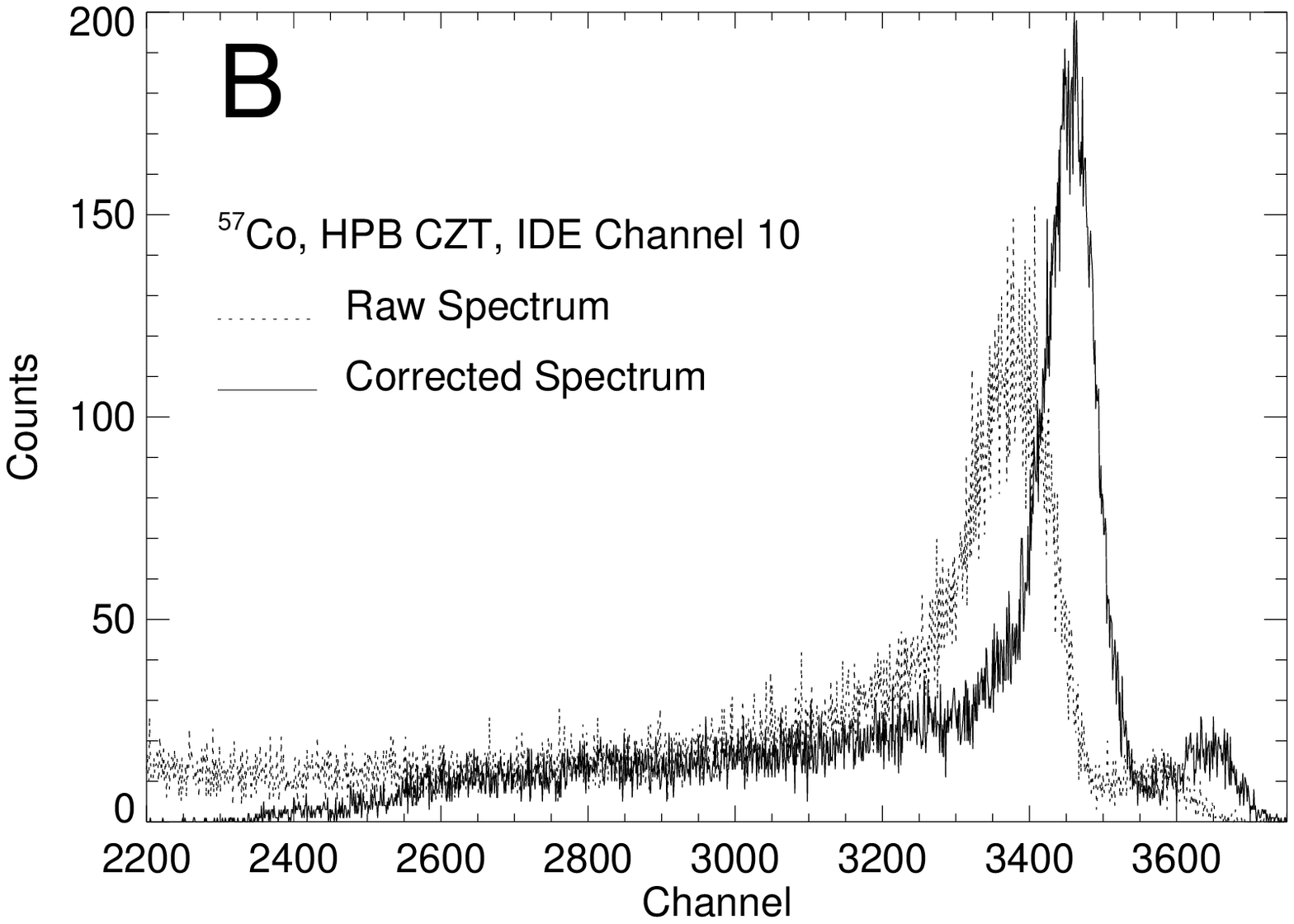,height=2.1in,width=3.2in}
\end{tabular}
\end{center}
\caption[initial]
{\label{fig:improve}
A) Improvement in an IMARAD pixel (IDE channel 26) $^{57}$Co spectrum
due to 
data post-processing.  The energy resolution improves from 15.6\% to
10.2\%. B) Improvement in an eV Products pixel (IDE channel 10)
spectrum due to post-processing.  The energy resolution improves from
8.9\% to 5.2\%. 
}
\end{figure}

The improvement in our $^{57}$Co spectra due to our data processing is
shown in 
Figure~\ref{fig:improve}.  
Both an IMARAD HB CZT pixel and an eV Products HPB CZT pixel are
shown.  The raw spectrum is shown by the dotted line, and the
corrected spectrum by the solid line.  In both cases the energy
resolution $E_{res}$ improves dramatically: from 15.6\% to 10.2\% on
the IMARAD 
pixel and from 8.9\% to 5.2\% on the eV Products pixel.  Charge
sharing between pixels is clearly an important effect even with large
2.35 mm pixels.

\begin{figure}
\begin{center}
\begin{tabular}{c}
\psfig{figure=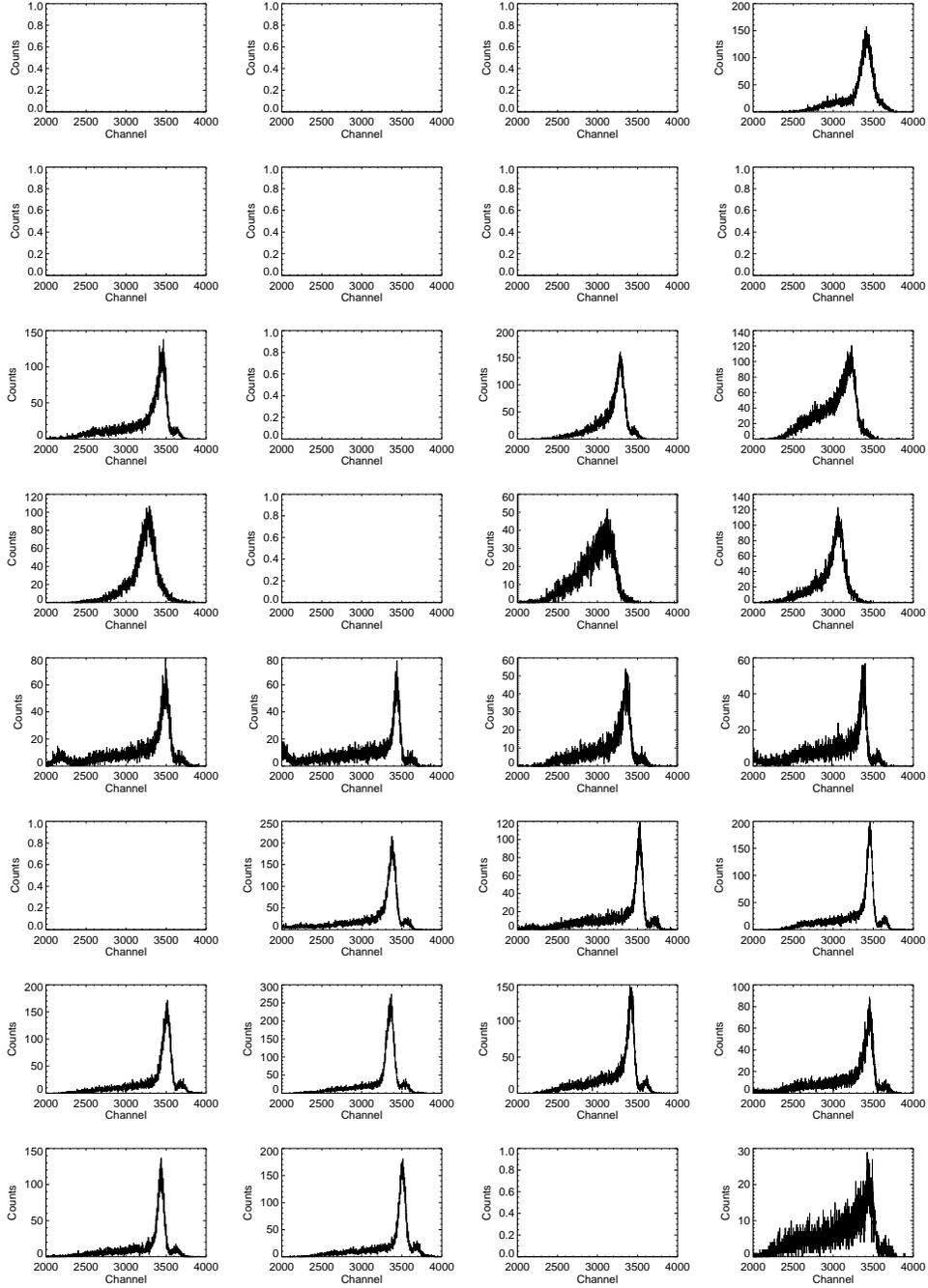,height=18cm} 
\end{tabular}
\end{center}
\caption[initial] 
{ \label{fig:fullflood}        
Corrected $^{57}$Co spectra in all working pixels of the imaging CZT
experiment.  The IMARAD detector makes up the top $4 \times 4$ plots
and the eV Products detector the bottom $4 \times 4$ plots.  The plots
are arranged so that the VA-TA ASIC sits below the figure.  The noisy
channels on the IMARAD detector, at the top of the figure, are on the
far side from ASIC and so have the longest lead lengths.  The good
IMARAD channels happen to correspond to bad pixels due to material
defects (see Figure~\ref{fig:preflip}) with the exception of one good
pixel on the left edge.  $\bar{E}_{res} = (16 \pm
9)$\% for the IMARAD detector and $\bar{E}_{res} = (7.3 \pm 2.8)$\%
for the eV Products detector.
} 
\end{figure}

The corrected $^{57}$Co spectra for all working pixels are shown in
Figure~\ref{fig:fullflood}.  
The spectra are arranged so that the ASIC would sit at the
bottom of the figure.  It is apparent that the noisy IMARAD pixels are
located on the far side of the detector from the ASIC and so suffer
from long lead lengths.  Our final intended detector-ASIC coupling
scheme would
eliminate this problem by placing the ASIC inputs directly below the
pixels. 
The remaining IMARAD channels happen to
correspond to those pixels with poor material qualities, as shown in
Figure~\ref{fig:preflip}, with the exception of one good pixel on the
left edge.  Thus the average of the energy resolution
values on the
IMARAD detector, with its standard deviation, is $\bar{E}_{res} = (16
\pm 9)$\%, comparable to 
scintillator detectors.  

The spectra on the eV Products side of the detector are quite good.
The variations in the number of counts are due partly to material
variations and partly to uneven illumination from the narrow beam of
our $^{57}$Co source.  The average energy resolution on the eV
Products detector is $\bar{E}_{res} = (7.3 \pm 2.8)$\%; this number
drops to $\bar{E}_{res} = (6.7 \pm 1.2)$\% if the lower right pixel,
shown in Figure~\ref{fig:preflip} to be bad, is excluded.  These
spectra are the best that we have seen published from a CZT detector,
read out by an ASIC, in a fully flight-ready configuration.

\section{Detector Simulations and Predicted Background}
\label{sec:sim}

We have carried out preliminary simulations of the tiled CZT detector
to determine its response and predict the expected flight background
spectrum and uniformity.  These simulations were performed in two parts.
First, the propagation of X-ray photons in the experiment materials and
their interaction in the detector were simulated using MGEANT\cite{mgeant}.  
MGEANT is an improved user interface to the CERN Program Library Monte Carlo
simulation package GEANT, allowing the easy specification of detector 
materials and geometry as well as the input particle spectrum and 
distribution.  For every event interacting in the CZT the energy deposited
and the location were recorded.  All relevant physical processes, including
photoelectric absorption, Compton scattering, and pair production, were
included. 

Second, the charge transport and signal induction on the pixels were
calculated.  We followed a procedure similar to that given by Kalemci
et al\cite{kalemci}.  First the electric field $\vec{E}$ and weighting
potential\cite{ramo} $W_{pot}$ were calculated for our pixel geometry
using the 
commercial electrostatics program ES3.  The electric field calculation
assumed 500 V on the cathode and 0 V on the pixels, while the
weighting potential calculation assumed a potential of 1 on the pixel
of interest and 0 on all other electrodes.  The values of $\vec{E}$ and
$W_{pot}$ were calculated on a 5 $\mu$m grid.  We used only a two-dimensional
slice through the center of an inner pixel for the present
simulations; the interaction positions given by MGEANT were therefore
collapsed into two dimensions.  Charges were then propagated along the
grid one step at a time, according to the components $E_z$ and $E_x$
of the electric field.  For each step the charge $Q_{ind}$ induced on
the pixel 
is given by the incremental change in the weighting
potential\cite{ramo}:
\begin{equation}
Q_{ind} = Q_{drift} \times \Delta W_{pot},
\end{equation}
where $Q_{drift}$ is the total of the charges drifting along
$\vec{E}$.  The number of drifting charges will gradually decrease due
to trapping.  We employed a simple exponential expression to describe
the amount of charge $Q_{drift}^{\prime}$ that arrives at each point
in the grid due to trapping at the previous point\cite{kalemci}:
\begin{equation}
Q_{drift}^{\prime} = Q_{drift} e^{-L_{drift}/L_{trap}},
\end{equation}
where $L_{drift}$ is the distance drifted between grid points and the
trapping length $L_{trap}$ is given by
\begin{equation}
L_{trap} = \mu \tau (E_x^2 + E_z^2)^{1/2},
\end{equation}
where $\mu \tau$ is the product of the charge carrier mobility and
lifetime.  The charge clouds were assumed to be points with no
diffusion.  Each cloud was followed until it reached an electrode and
the total induced charge for that event was tabulated.  The resulting
histogram of detector pulse heights was finally convolved with a
gaussian to simulate electronic noise.

We first used MGEANT to illuminate the CZT detectors with a simulated
$^{57}$Co spectrum to compare with our calibration data.  
Both recorded and simulated spectra were fit with a combination of a
gaussian photopeak and an exponential tail to determine
$E_{res}$ and the photopeak efficiency\cite{narita98}.
The photopeak efficiency is defined as the ratio of the photopeak
counts to the total counts within a
range around the gaussian center from $-4\sigma$ to $+2.35\sigma$.
The electron
mobility-lifetime product $\mu_e \tau_e$ has been well-measured in HPB
CZT\cite{he}, and we fixed this quantity at a typical value of $3
\times 10^{-3}$ cm$^2$ V$^{-1}$.  (Our previous work indicates similar
$\mu_e \tau_e$ values in IMARAD material\cite{narita99}.)  The hole
mobility-lifetime 
product $\mu_h \tau_h$ and convolving gaussian width $\sigma$ at 122
keV were 
then adjusted to generate the best match in energy resolution and
photopeak efficiency to the recorded data.  
We found the best
agreement for values of $\mu_h \tau_h = 10^{-6}$ cm$^2$ V$^{-1}$ and
$\sigma = 2.5$ keV.  This value of $\mu_h \tau_h$ is somewhat lower
than typical values found in the literature\cite{eisen}.
\begin{figure}
\begin{center}
\begin{tabular}{lr}
\psfig{figure=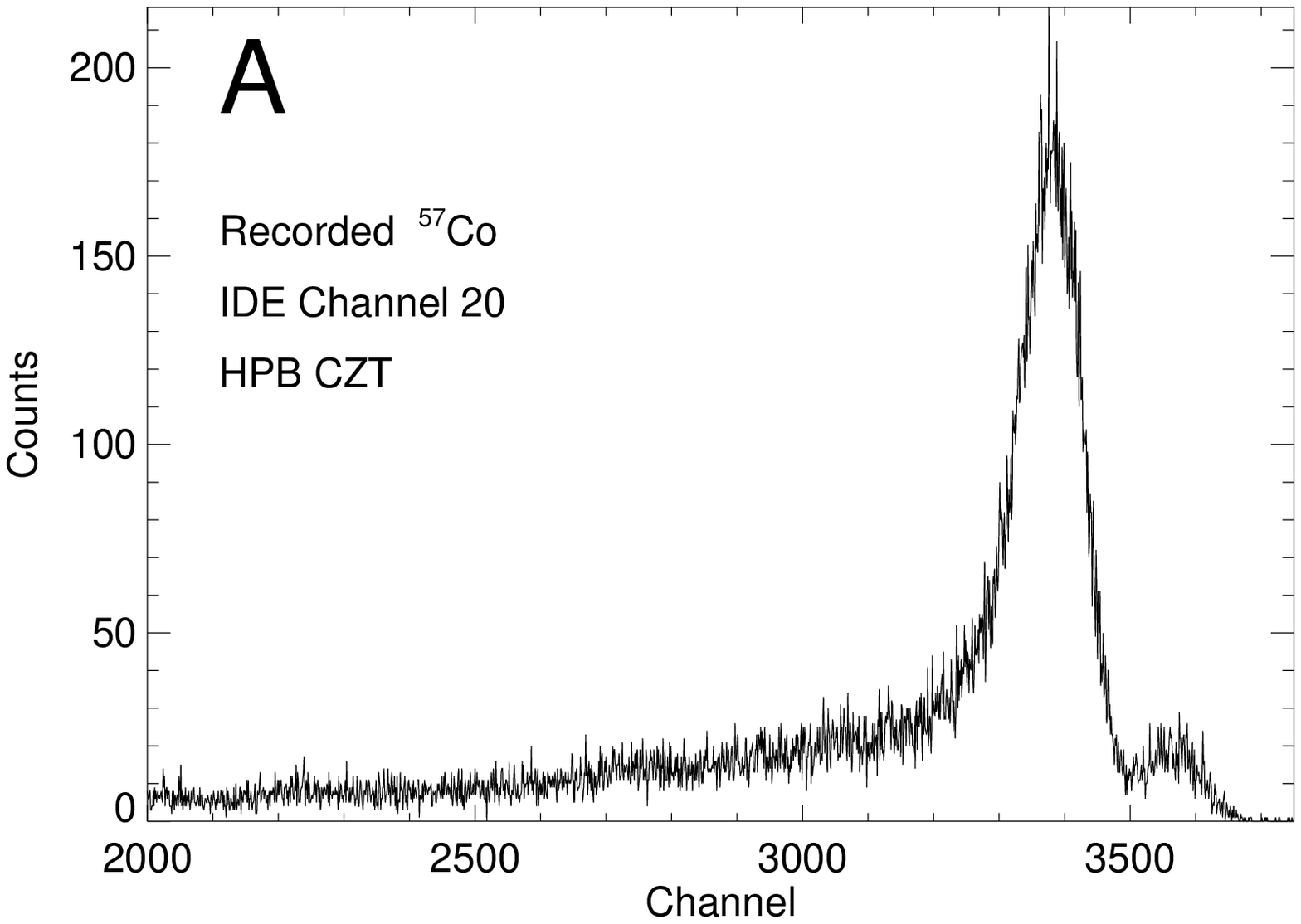,height=2.1in,width=3.2in}
& \psfig{figure=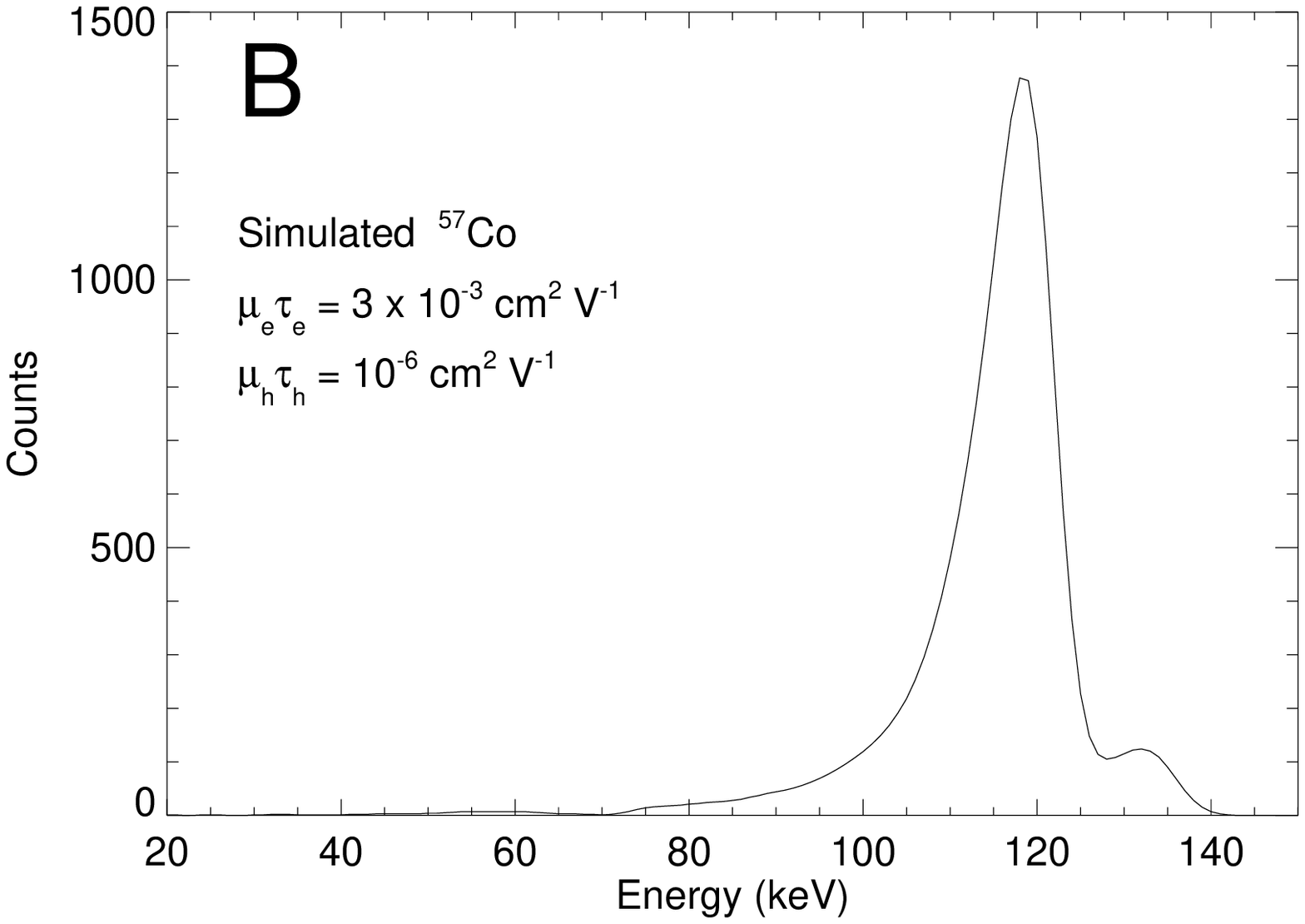,height=2.1in,width=3.2in}
\end{tabular}
\end{center}
\caption[initial]
{\label{fig:compare_sim}
A) Recorded $^{57}$Co spectrum from IDE channel 20, an inner pixel of
the eV Products detector. Here $E_{res} = 7.3$\% and photopeak
efficiency = 84\%.  B) Simulated $^{57}$Co spectrum using
$\mu_e \tau_e = 3 \times 10^{-3}$ cm$^2$ V$^{-1}$, $\mu_h \tau_h =
10^{-6}$ cm$^2$ V$^{-1}$, and $\sigma = 2.5$ keV.  This corresponds to
an electronic noise of $\sim 540$ $e^{-}$.  The simulation mimics
the photopeak quite well, but does not reproduce the tail completely.
Here $E_{res} = 7.6$\% and photopeak efficiency = 75\%.
}
\end{figure}
In Figure~\ref{fig:compare_sim} we show the corrected $^{57}$Co
spectrum recorded from an inner pixel of the eV Products HPB detector
together 
with our simulated $^{57}$Co spectrum using these parameters.  The
recorded spectrum has an energy resolution of 7.3\% and a photopeak
efficiency of 84\%.  The simulated spectrum has an energy resolution
of 7.6\% and a photopeak efficiency of 75\%.  It is obvious that the
photopeak is accurately reproduced, but not the tail, which in the
real data extends to lower energies with a flatter slope.  This could
be due to additional trapping in the corners of the pixel, which is
not included in our two-dimensional model.  Assuming that the average
electron-hole pair creation energy is 4.6 eV\cite{muller}, the 2.5 keV
gaussian width corresponds to an equivalent noise of $\sim 540$ $e^-$,
compared to the 300 $e^-$ noise that the VA-TA system should
provide under ideal conditions.  The difference is likely due to
capacitance effects from the long lead lengths.

Using the parameters given above for the CZT detector response, we
have simulated the expected background spectrum in one inner pixel at our
planned balloon float altitude, corresponding to 3.5 g cm$^{-2}$
residual atmosphere.  A model of the aluminum pressure vessel and
internal hardware was created using MGEANT, along with the Pb/Sn/Cu
passive collimator and shield.  The pressure vessel was irradiated
with the measured atmospheric gamma-ray spectrum between 20 keV and 10
MeV\cite{gehrels}, assumed here to be isotropic.  
\begin{figure}[t] 
\begin{minipage}[t]{3.3in}
\psfig{file=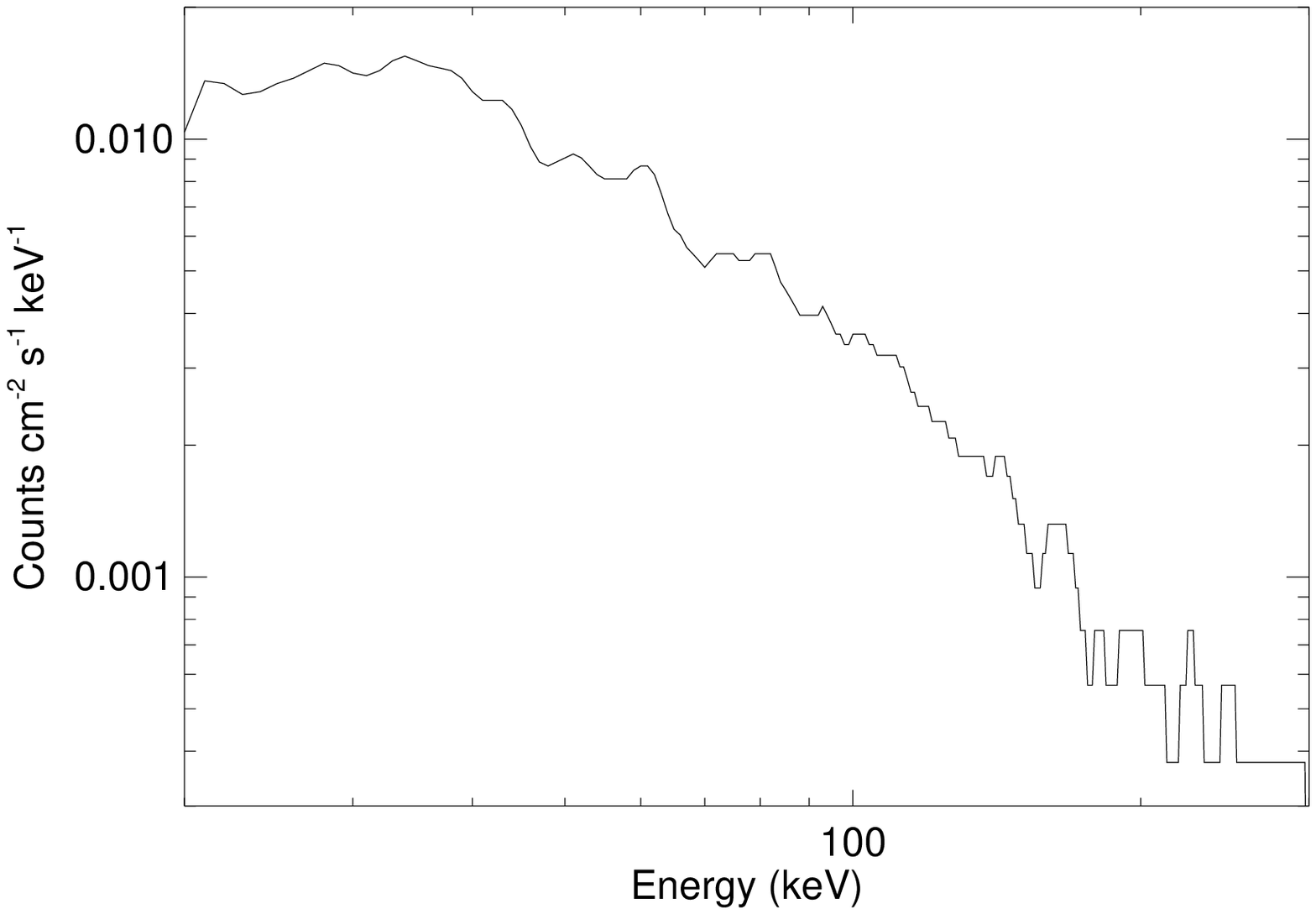,height=2.1in,width=3.2in}
\caption{Predicted background at 3.5 g cm$^{-2}$ residual atmosphere
for the imaging CZT detector.  Only photon interactions are included
and the plastic scintillator is assumed to veto all local gamma
production.  The background at 100 keV is $\sim 3.6 \times 10^{-3}$
cts cm$^{-2}$ s$^{-1}$ keV$^{-1}$. 
}
\label{fig:background}
\end{minipage}
\hspace*{0.2in}
\begin{minipage}[t]{3.3in}
\psfig{file=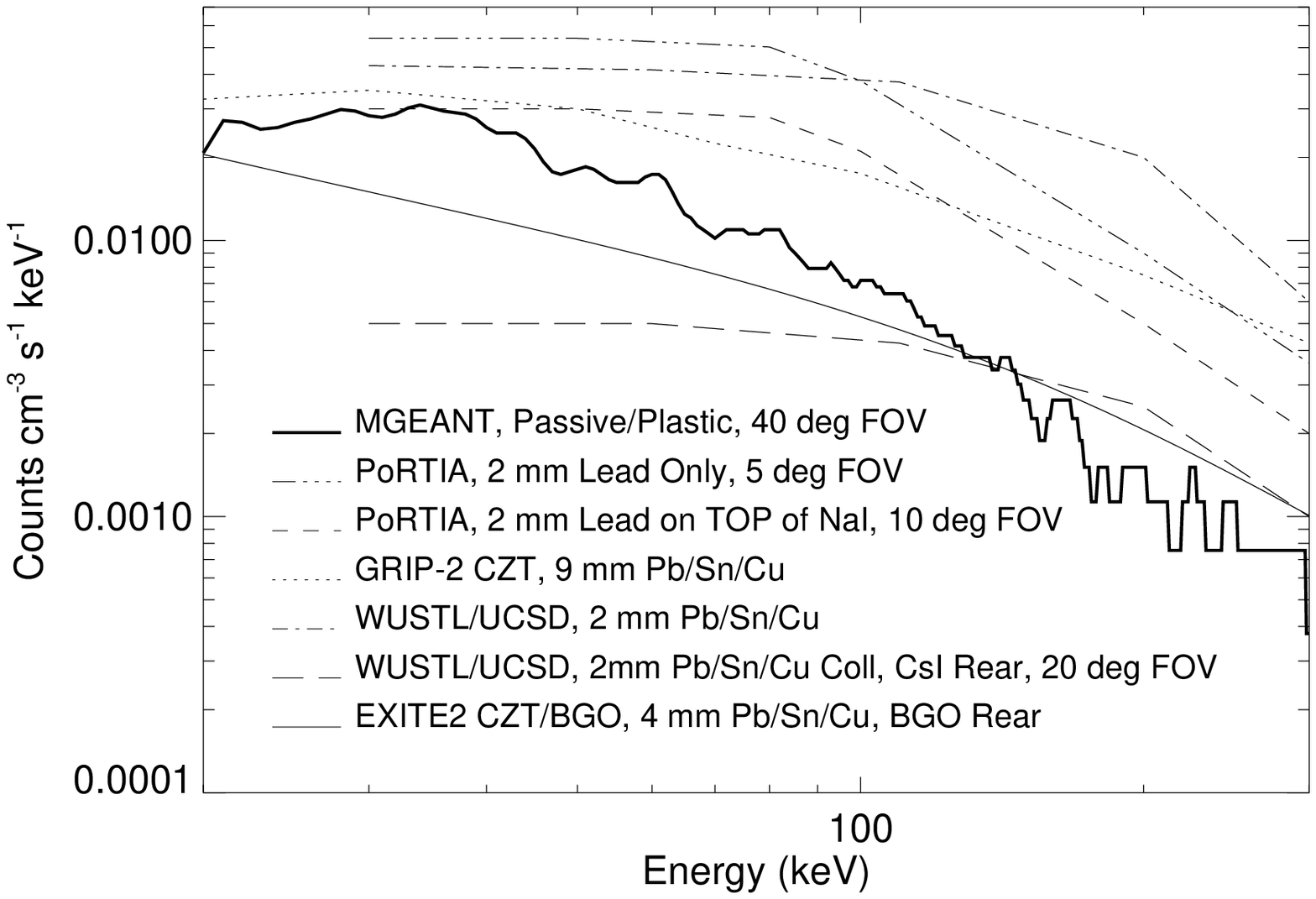,height=2.1in,width=3.2in}
\caption{Comparison of our predicted background with previous
measurements, plotted per detector volume to account for different
detector thicknesses.  The predicted level is lower than that in
totally passive experiments, and comparable to that in experiments
employing some active shielding.}
\label{fig:background_vol}
\end{minipage}
\end{figure}
The resulting spectrum in one pixel of the CZT is shown in
Figure~\ref{fig:background}.  This spectrum represents a best case,
assuming that the plastic scintillator shields veto all local
gamma-ray production from charged particles.  The spectrum has been
convolved with a gaussian whose width varies according to $\sigma =
2.5(E/122 {\rm keV})^{1/2}$ keV.  The background level at 100 keV is
$\sim 3.6 \times 10^{-3}$ cts cm$^{-2}$ s$^{-1}$ keV$^{-1}$.

For comparison, we have plotted in Figure~\ref{fig:background_vol} the
predicted background together with the results of several CZT balloon
experiments.   Since most previous balloon flights have involved 2 mm
thick detectors, we plot the backgrounds per detector volume.  Several
experiments used only passive shielding: the first flight of the
Goddard Space Flight Center experiment PoRTIA\cite{parsons96}, with 2
mm Pb shielding around a 25.4 mm $\times$ 25.4 mm $\times$ 1.9 mm
detector and a $5^{\circ}$ field of view, the Caltech CZT detector
flown on the GRIP-2 payload\cite{harrison98}, with 9 mm of graded Pb,
Sn, and Cu completely shielding a 10 mm $\times$ 10 mm $\times$ 2 mm
detector, and the first flight of the Washington University, St. Louis
and University of California-San Diego (WUSTL/UCSD) cross-strip
detector\cite{slavis98}, with 2 mm of graded Pb, Sn, and Cu completely
surrounding a 12 mm $\times$ 12 mm $\times$ 2 mm employing orthogonal
strips electrodes for readout.  The backgrounds recorded by these
experiments are all higher than that expected from our tiled CZT
detectors, due mainly to the local production of gamma-rays in the
passive shielding by cosmic-ray interactions.  Other previous
experiments have used partial active shielding: the second flight of
PoRTIA\cite{parsons96}, in which the detector was placed on top of a
large NaI crystal (now with a 10$^{\circ}$ field of
view), the second flight of the WUSTL/UCSD detector\cite{slavis99},
which employed a 2 mm Pb/Sn/Cu collimator with a 20$^{\circ}$ field of
view and a rear CsI shield, and our previous CZT experiment flown on
EXITE2\cite{bloser98}, which completely enclosed a 10 mm $\times$ 10
mm $\times$ 2 mm CZT detector in 4 mm of Pb/Sn/Cu in the front and a
thick BGO shield in the rear.  The active/passive WUSTL/UCSD and
EXITE2 CZT/BGO background levels are comparable to the predicted tiled
detector background, though lower at low energies due to their smaller
fields of view.  It is not clear why the active/passive PoRTIA
spectrum is still so high; perhaps the solid angle subtended by the
NaI shield was not sufficient to veto local gammas.  The previous
experiments clearly demonstrate that at least partial active shielding
is required to 
achieve low background levels.  The second WUSTL/USCD flight
showed in addition that most of the background reduction was achieved
with the active shield threshold set near 10 MeV, appropriate for
rejecting charged particles\cite{slavis99}.  This implies that CZT internal
gamma-ray background is of secondary importance, and that our
plastic shielding scheme should indeed produce the background level
predicted.  We will be able to test this hypothesis on our September
flight: we are re-flying the CZT/BGO detector alongside the tiled
detectors to compare shielding configurations\cite{bloser99}, and will
take data with 
the BGO threshold set near 50 keV (to veto gamma-rays) and near 1 MeV
(to veto charged particles only) to see if the gamma-ray rejection
is needed. 

\begin{figure}
\begin{center}
\begin{tabular}{c}
\psfig{figure=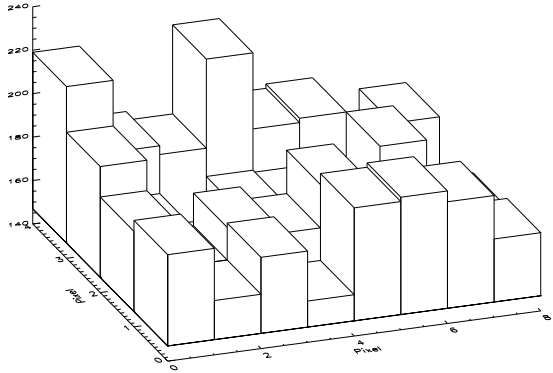,height=6cm}
\end{tabular}
\end{center}
\caption[initial]
{\label{fig:uniformity}
Predicted uniformity of the flight background from MGEANT
simulations.  Plotted are the total counts recorded in each pixel.
The edge pixels have generally higher background rates, as expected,
though the significance is low.
}
\end{figure}
The expected uniformity of the flight background as given by the
MGEANT simulation is shown in Figure~\ref{fig:uniformity}.  We plot
here the total counts recorded in each pixel.  As might be expected,
the background level in the inner pixels is generally lower due to
shielding from the outer pixels; this difference is only about 10\% on
average however, and the significance is only $\sim 1\sigma$.

\section{Conclusions and Future Work}
\label{sec:conc}

We have constructed an imaging CZT detector that demonstrates a number
of the techniques required for a hard X-ray survey telescope such as
that needed for EXIST.  These techniques include the use of thick
pixellated detectors with
blocking contacts, flip-chip mounting, tiled detector geometry, ASIC
readout, and shielding.  The spectra recorded by the eV Products
detector, on the short-lead-length side of the detector array, are the
best we have seen for a CZT detector read out by an ASIC in a
flight-ready configuration.
Our immediate goal is to finally fly this
experiment on the EXITE2/HERO balloon payload from Ft. Sumner, NM in
September 2000.  Our future work will include the design of an
optimized ASIC and coupling scheme to minimize lead lengths (ideally
coupling the ASIC directly to the CZT pixels), the improvement of our
flip-chip processing, and the refinement of our detector simulations
to include three-dimensions and more complicated trapping-detrapping
effects.  Using what we learn from the September flight we will begin
the development of EXIST-LITE, a proposed ultra-long-duration balloon
hard X-ray telescope employing $\sim 1$ m$^2$ of CZT to conduct a
preliminary survey of half the sky.

\acknowledgments     
 
We thank S. Sansone and L. Knowles for machining work and 
J. Gomes, G. Nystrom, and F. Licata for engineering assistance.  
We also thank K. Shah and P. Bennett at RMD, Inc. for their assistance
in fabricating the detectors and G. Riley at HyComp, Inc. for his help
in the flip-chip processing.  We
especially thank B. Sundal at IDE AS for extensive help with the
VA-DAQ system. 
This work was supported in 
part by NASA grants NAG5-5103 and NAG5-5209.  P. Bloser acknowledges
partial support from NASA GSRP grant NGT5-50020.


\bibliography{czt2_paper}   
\bibliographystyle{spiebib}   
 
\end{document}